\documentclass[aps,prb, amsmath,amssymb,floatfix,12pt]{revtex4-1}
\usepackage{tabularx}
\usepackage{bm}
\usepackage{euscript}
\usepackage{graphicx}
\usepackage{color}
\usepackage{amsfonts}
\usepackage{exscale}
\usepackage{amsbsy}
\usepackage{subfigure}
\usepackage{textcomp}
\usepackage{comment}

\usepackage{hyperref}

\numberwithin{equation}{section}

\newcommand{\be}{\begin{equation}}
\newcommand{\ee}{\end{equation}}
\newcommand{\bea}{\begin{eqnarray}}
\newcommand{\eea}{\end{eqnarray}}

\begin{document}

\title{Emergent particle-hole symmetry in the half-filled Landau level }
\author{Michael Mulligan$^1$, S. Raghu$^{1,2}$, and Matthew P. A. Fisher$^3$}
\affiliation{$^{1}$Stanford Institute for Theoretical Physics, Stanford University, Stanford, California 94305, USA}
\affiliation{$^2$SLAC National Accelerator Laboratory, 2575 Sand Hill Road, Menlo Park, CA 94025, USA}
\affiliation{$^3$Department of Physics, University of California, Santa Barbara, California 93106, USA}

\date{\today}
\begin{abstract}
We provide an effective description of a particle-hole symmetric state of electrons in a half-filled Landau level, starting from the traditional approach pioneered by Halperin, Lee and Read.
Specifically, we study a system consisting of alternating  quasi-one-dimensional strips of composite Fermi liquid (CFL) and composite hole liquid (CHL), both of which break particle-hole symmetry. When the CFL and CHL strips are identical in size, the resulting state is manifestly invariant under the combined action of a particle-hole transformation with respect to a single Landau level (which interchanges the CFL and CHL)  and translation by one unit, equal to the strip width, in the direction transverse to the strips. 
At distances long compared to the strip width, we demonstrate that the system is described by a  Dirac fermion coupled to an emergent gauge field, with an anti-unitary particle-hole symmetry, as recently proposed by Son.
\end{abstract}

\maketitle

\tableofcontents

\section{Introduction}

An influential way of understanding much of the physics of the quantum Hall regime invokes the idea of composite fermions [\onlinecite{jainCF}].  
In essence, a composite fermion can be viewed as an electron (or hole) bound to several flux quanta. The challenging problem of electrons strongly interacting with one another in a partially filled Landau level can be recast in terms of (hopefully) simpler effective descriptions involving composite fermions.  
While ultimately an inspired guess, the composite fermion perspective nevertheless leads to an elegant understanding of all observed quantum Hall states.  Furthermore, theories based on composite fermions have led to several spectacular predictions [\onlinecite{jainCF}] - including that of a gapless metallic ground state in a half-filled ($\nu=1/2$) Landau level - which have been experimentally observed [\onlinecite{Jiang1989transportanomalies, willett1990, willett1993}].   


In a pioneering study, Halperin, Lee and Read [\onlinecite{halperin1993}] (as well as Kalmeyer and Zhang [\onlinecite{kalmeyer1992}]) suggested that this gapless state at half-filling ought to be a metallic state of composite fermions - a {\it composite Fermi liquid} (CFL), in zero background magnetic field.  
This approach is a simple effective field theory in which electrons bound to two flux quanta form a filled Fermi sea.  
The effective description involves composite fermion ``quasiparticles" that are  coupled to a Chern-Simons gauge field - which implements flux attachment - and the resulting system is strongly coupled.  Despite such difficulties the CFL theory has led to a widely accepted qualitative understanding of the half-filled Landau level, and its instabilities to incompressible states as are seen in the second Landau level at $\nu=5/2$ [\onlinecite{willett1987}].  


Although the CFL theory motivates a physically-compelling mean-field description of the half-filled Landau level, it suffers from a drawback: it strongly breaks particle-hole (PH) symmetry, which is a property of the half-filled lowest Landau level Hamiltonian in the limit of vanishing Landau level mixing, in apparent contradiction with electrical transport experiments [\onlinecite{Shahar1995, Shahar1996, Wong1996}].\footnote{Our statement regarding the observation of particle-hole symmetry at half-filling assumes adiabatic continuity of the measurements of strongly disordered samples to those with less disorder.}  
Indeed, as suggested in [\onlinecite{kivelson1997}], in order that the electrical Hall conductivity $\sigma_{xy} = {1 \over 2} {e^2 \over h}$ -- the value required by PH symmetry at half-filling -- the composite fermions must exhibit a large Hall conductivity $\sigma_{xy}^{(cf)} = - {1 \over 2} {e^2 \over h}$.\footnote{An alternative possibility requires the composite fermion resistivity $\rho_{xx}^{(cf)}$ to be vanishingly small, however, in experimental systems, the measured $\rho_{xx} = \rho_{xx}^{(cf)} \sim (1/10 - 1) {h \over e^2}$, thereby contradicting the approximate PH symmetry.}  
Since the composite fermions ``feel" zero magnetic field on average at half-filling, one instead expects $\sigma_{xy}^{(cf)} = 0$ and concludes that PH symmetry must be broken. 


Remarkable recent experiments [\onlinecite{Kamburov2014, KamburovMueed2014,LiuDengWaltz}] highlighting the importance of PH symmetry (and the PH transformation, more generally) in the half-filled lowest Landau level led to the study [\onlinecite{BMF2015}] of an alternate construction of the $\nu=1/2$ state, the {\it composite hole liquid} (CHL), which consists of holes of a filled Landau level, at half-filling, attached to two flux quanta.\footnote{See \cite{BalramRifmmodeCsabaJain2015} for an alternative interpretation of the experiments in \cite{Kamburov2014, KamburovMueed2014,LiuDengWaltz}.}
In a theory that admits a spontaneous breaking of PH symmetry, the CHL and CFL may be thought of as two degenerate ground states, one of which is spontaneously chosen, in analogy with the up and down spin states of an Ising model.  Alternatively, a first order transition may also occur between these two distinct states.




Our goal here is to understand how PH symmetry might be restored, starting from the symmtry-broken CFL and CHL formulations.
At first glance, such a task appears to be hopeless: it is loosely analogous to constructing a description of the critical point of the Ising model in terms of the ordered phases that describe the physics far away from criticality.  
Nevertheless, by considering an inhomogeneous configuration built from both the CFL and CHL, we will show that it is possible to restore PH symmetry on long length scales.
In analogy to the Ising model, it is as though we attempt to construct a critical state by studying domains of up and down spin configurations.  With the help of hindsight, we know that such a construction can  indeed restore the symmetry: in the case of the Ising model, fermionic modes associated with the domain walls are deconfined at the critical point, and provide a description of the emergent critical behavior.


More specifically, we describe the PH symmetric $\nu=1/2$ state by considering a two-dimensional (2D) spatial plane patterned with CFL and CHL strips (Fig. \ref{patternedplane}), where domain walls between the two regions possess new emergent fermions that will ultimately represent the low-energy degrees of freedom of a PH symmetric half-filled Landau level.  As we show below, this strategy leads to a successful description of the PH symmetric half-filled LL.  
Our configuration enjoys the combined symmetry of a PH transformation and translation by one ``unit," equal to the strip width, in the direction transverse to the strips.  In the long-wavelength limit, {\it i.e.} for lengths much larger than the width of these strips, the resulting system at low energies consists of an electrically-neutral composite Dirac particle coupled to an emergent U(1) gauge field (analogous to [\onlinecite{MrossEssinAlicea2015}]), with an anti-unitary implementation of the PH symmetry, precisely of the form conjectured by Son in a remarkable recent paper [\onlinecite{Son2015}].  
Recent work [\onlinecite{WangSenthilfirst2015, maxashvin2015, KMTW2015, WangSenthilsecond2016, Geraedtsetal2015, MurthyShankar2016halfull, mrossaliceamotrunich2015}] has sought to confirm and explore the duality proposed by Son.  
While our construction shows that there is in principle an RG flow between the CFL/CHL states and the composite Dirac liquid, other possibilities, including first order transitions between the CFL and CHL can also occur, but will not be considered here.   

The paper is organized as follows.
In Sec. \ref{compositefermionsboundary}, we review the composite Fermi and composite hole theories and describe the action of the particle-hole transformation. 
In Sec. \ref{sec:p-hsymemergence}, we show how a particle-hole symmetric composite Fermi liquid emerges at long wavelengths from the patterned plane configuration.
We conclude and mention a few possible directions of future work in Sec. \ref{conclusion}.

\section{Composite electrons and holes}
\label{compositefermionsboundary}
\subsection{Bulk description}

The traditional approach to the effective description of the half-filled lowest Landau level (LLL) involves the composite Fermi liquid lagrangian, ${\cal L}_{\rm CFL} = {\cal L}_{f} + {\cal L}_{\rm gauge} + {\cal L}_{\rm int}$ (see Refs. \cite{jainCF, simon1998, Fradkinbook} for reviews):
\begin{align}
\label{bulkcompositefermionlagrangian}
{\cal L}_{f} & = f^\dagger \Big(i\partial_t + (a_t + A_t) + {1 \over 2 m_f} (\partial_j - i (a_j + A_j))^2 \Big) f, \cr
{\cal L}_{\rm gauge} & = {1 \over 2} {1 \over 4 \pi} \epsilon_{\mu \nu \rho} a_\mu \partial_\nu a_\rho, \cr
{\cal L}_{\rm int} & = - {1 \over 2} \int d^2 \mathbf{r}' \Big(f^\dagger f(\mathbf{r}) - n_{f} \Big) U_{\mathbf{r}, \mathbf{r}'} \Big(f^\dagger f(\mathbf{r}') - n_{f} \Big).
\end{align}
$f$ is the destruction operator of a composite electron with effective mass $m_f$, $a_\mu$ is an emergent gauge field, $A_\mu$ represents the electromagnetic gauge field with non-zero average magnetic flux $\langle \partial_x A_y - \partial_y A_x \rangle = B > 0$  
(the indices $\mu, \nu, \rho \in \{t,x,y\}$, the totally anti-symmetric symbol is defined by fixing $\epsilon_{txy} = 1$, the two spatial coordinates $\mathbf{r} = (x,y)$ and $\mathbf{r}' = (x',y')$, and both gauge charges have unit magnitude).  
The average density of composite electrons $\langle f^\dagger f \rangle \equiv n_f$.

A partially-filled LLL can equally well be described in terms of holes of the filled Landau level.
Thus, we might instead describe the half-filled LLL using the composite hole liquid lagrangian, ${\cal L}_{\rm CHL} = 
{\cal L}_h + {\cal L}_{\rm gauge} + {\cal L}_{\rm int}$\cite{BMF2015},
\begin{align}
\label{bulkcompositeholelagrangian}
{\cal L}_{h} & = h^\dagger \Big(i\partial_t + (b_t - A_t) + {1 \over 2 m_h} (\partial_j - i (b_j - A_j))^2 \Big) h, \cr
{\cal L}_{\rm gauge} & = - {1 \over 2} {1 \over 4 \pi} \epsilon_{\mu \nu \rho} b_\mu \partial_\nu b_\rho + {1 \over 4 \pi} \epsilon_{\mu \nu \rho} A_\mu \partial_\nu A_\rho, \cr
{\cal L}_{\rm int} & = - {1 \over 2} \int d^2 \mathbf{r}' \Big(h^\dagger h(\mathbf{r}) - n_{h} \Big) \tilde{U}_{\mathbf{r}, \mathbf{r}'} \Big(h^\dagger h(\mathbf{r}') - n_{h} \Big),
\end{align}
where now, $h$ is the destruction operator of a composite hole with effective mass $m_h$, and $b_{\mu}$ is an emergent gauge field.  The second term in ${\cal L}_{\rm gauge}$ describes the filled Landau level vacuum, which leads to an integer Hall conductance.  Note that the electromagnetic charge of the composite holes is equal in magnitude but opposite in sign to that of the composite electrons.  

Before we proceed with an explicit construction of a PH symmetric half-filled Landau level, we require that the two Lagrangians above yield consistent predictions for the electronic properties of the $\nu=1/2$ state.  For instance, the physical electromagnetic charge density follows from differentiating both ${\cal L}_{\rm CFL}$ and ${\cal L}_{\rm CHL}$ with respect to $A_t$.  Equating the resulting operators, we find:
\begin{equation}
f^{\dagger} f + h^{\dagger} h = \frac{B}{2 \pi}.
\end{equation} 

At a heuristic level, ${\cal L}_{\rm CFL}$ and ${\cal L}_{\rm CHL}$ describe the dynamics of the particles that result from attaching two units of flux to the electrons or to the holes of the lowest Landau level.  
Away from half-filling, the system of interest may be described either in terms of composite electrons or composite holes: the two descriptions represent a different way of encoding the same state.  For instance, the Laughlin state at $\nu=1/3$ can be described either as an integer quantum Hall state of composite electrons, $\nu_f = 1$, or of composite holes with a different filling fraction $\nu_h = 2$.  
At half-filling, however, PH symmetry requires that {\it the two descriptions must be one and the same}.  This requirement is obscured by the fact that treated at mean-field level, both ${\cal L}_{\rm CFL}$ and ${\cal L}_{\rm CHL}$ manifestly break PH symmetry.  We will see how to overcome this difficulty below.  For now, we note that in order for ${\cal L}_{\rm CFL}$ and ${\cal L}_{\rm CHL}$ to map into one another under a PH transformation, the couplings must satisfy $m_f = m_h$ and $U(x) = \tilde U(x)$.  In addition, we require that the anti-unitary PH transformation be implemented using the following three steps.  First, we transform
\begin{align}
(a_t, a_x, a_y) & \mapsto (a_t, - a_x, - a_y), \cr
(b_t, b_x, b_y) & \mapsto (b_t, - b_x, - b_y), \cr 
(A_t, A_x, A_y) & \mapsto (- A_t, A_x, A_y),
\end{align}
along with the mapping $(t, x, y) \mapsto (-t, x, y)$ and $i \mapsto - i$.\footnote{In terms of the charge-conjugation ${\cal C}$ and time-reversal ${\cal T}$ transformations, PH acts by ${\cal T}$ on the emergent gauge fields and coordinates, and by ${\cal CT}$ on the electromagnetic gauge field.}
Next, we shift 
${\cal L}_{\rm gauge} \rightarrow {\cal L}_{\rm gauge} + \frac{1}{4 \pi} \epsilon_{\mu \nu \rho} A_{\mu} \partial_{\mu} A_{\rho}$.\footnote{This shift is reminiscent of the shift of the three spatial dimensional quantum electrodynamic effective lagrangian by a term proportional $\epsilon_{\mu \nu \rho \tau} \partial_\mu A_{\nu} F^{\rho \tau}$ due to the (anomalous) transformation of the fermionic path integral measure under a chiral rotation.} 
Lastly, we take $f \mapsto h$, $h \mapsto f$, and relabel the two emergent gauge fields: $a_\mu \leftrightarrow b_\mu$.  In this way, ${\cal L}_{\rm CFL}$ and ${\cal L}_{\rm CHL}$ map into one another.  

\subsection{Boundary lagrangians}

If the CFL is placed in the lower half-plane $y < 0$ with a topologically trivial vacuum in the upper half-plane ($y>0$), gauge invariance requires the presence of boundary degrees of freedom at $y=0$ with lagrangian:
\begin{align}
\label{boundarycompositefermionlagrangian}
{\cal L}_{\partial {\rm CFL}} = {2 \over 4 \pi} \Big[ (\partial_t \phi_f + {1 \over 2} a_t) (\partial_x \phi_f + {1 \over 2} a_x) - v_{\phi_f} (\partial_x \phi_f + {1 \over 2} a_x)^2 + {1 \over 2} \epsilon_{\mu \nu y} V_\mu \partial_\nu \phi_f \Big] \delta(y=0).
\end{align}
Together, the boundary and bulk lagrangians of composite electrons are invariant under the gauge transformation,
\begin{align}
\label{cfgaugetrans}
a_\mu & \mapsto a_\mu + \partial_\mu \Lambda_a, \cr
f & \mapsto e^{i \Lambda_a} f, \cr
\phi_f & \mapsto \phi_f - {1 \over 2} \Lambda_a.
\end{align}
The first two terms in Eq. (\ref{boundarycompositefermionlagrangian}) are gauge invariant and describe the dynamics of the boundary modes.  The third is required to absorb the (anomalous) gauge transformation of the bulk Chern-Simons term in ${\cal L}_{\rm gauge}$ [\onlinecite{Elitzur89, wengaplessboundary, StoneIW,BMF2015}]. Since $f$ carries unit electromagnetic charge, it follows from above that the boundary operator $\psi_e = e^{2 i \phi_f} f$ carries unit electromagnetic charge but is neutral with respect to $a_{\mu}$.  Thus, we naturally identify $\psi_e$ with the electron destruction operator at a boundary.  

Likewise, if the CHL is placed in the lower half-plane (with the topologically-trivial vacuum in the upper half-plane), gauge invariance requires the addition of the boundary degrees of freedom:
\begin{align}
\label{boundarycompositeholelagrangian}
{\cal L}_{\partial {\rm CHL}} & = - {2 \over 4 \pi} \Big[ (\partial_t \phi_h - {1 \over 2} b_t) (\partial_x \phi_h - {1 \over 2} b_x) + v_{\phi_h} (\partial_x \phi_h - {1 \over 2} b_x)^2 - {1 \over 2} \epsilon_{\mu \nu y} b_\mu \partial_\nu \phi_h \Big] \delta(y=0) \cr
& + {1 \over 4 \pi} \Big[ \partial_x \phi_{\rm LL} (\partial_t - v_{\rm LL} \partial_x) \phi_{\rm LL} + 2 \epsilon_{\mu \nu y} A_\mu \partial_\nu \phi_{\rm LL} \Big] \delta(y=0),
\end{align}
so that the total system remains invariant under
\begin{align}
\label{chgaugetrans}
b_\mu & \mapsto b_\mu + \partial_\mu \Lambda_b, \cr
h & \mapsto e^{i \Lambda_b} h, \cr
\phi_h & \mapsto \phi_h + {1 \over 2} \Lambda_b.
\end{align}
In addition to the hole destruction operator, $\psi_h = e^{- 2 i \phi_h} h$ which has opposite electromagnetic charge as $\psi_e$ above, and is neutral with respect to $b_{\mu}$, there is a chiral boson field $\phi_{\rm LL}$ describing the edge mode of a filled Landau level: the corresponding edge electron destruction operator of the filled Landau level is $\psi_{\rm LL} \sim e^{i \phi_{\rm LL}}$.  

\begin{figure}[h!]
\centering
\includegraphics[width=.7\linewidth]{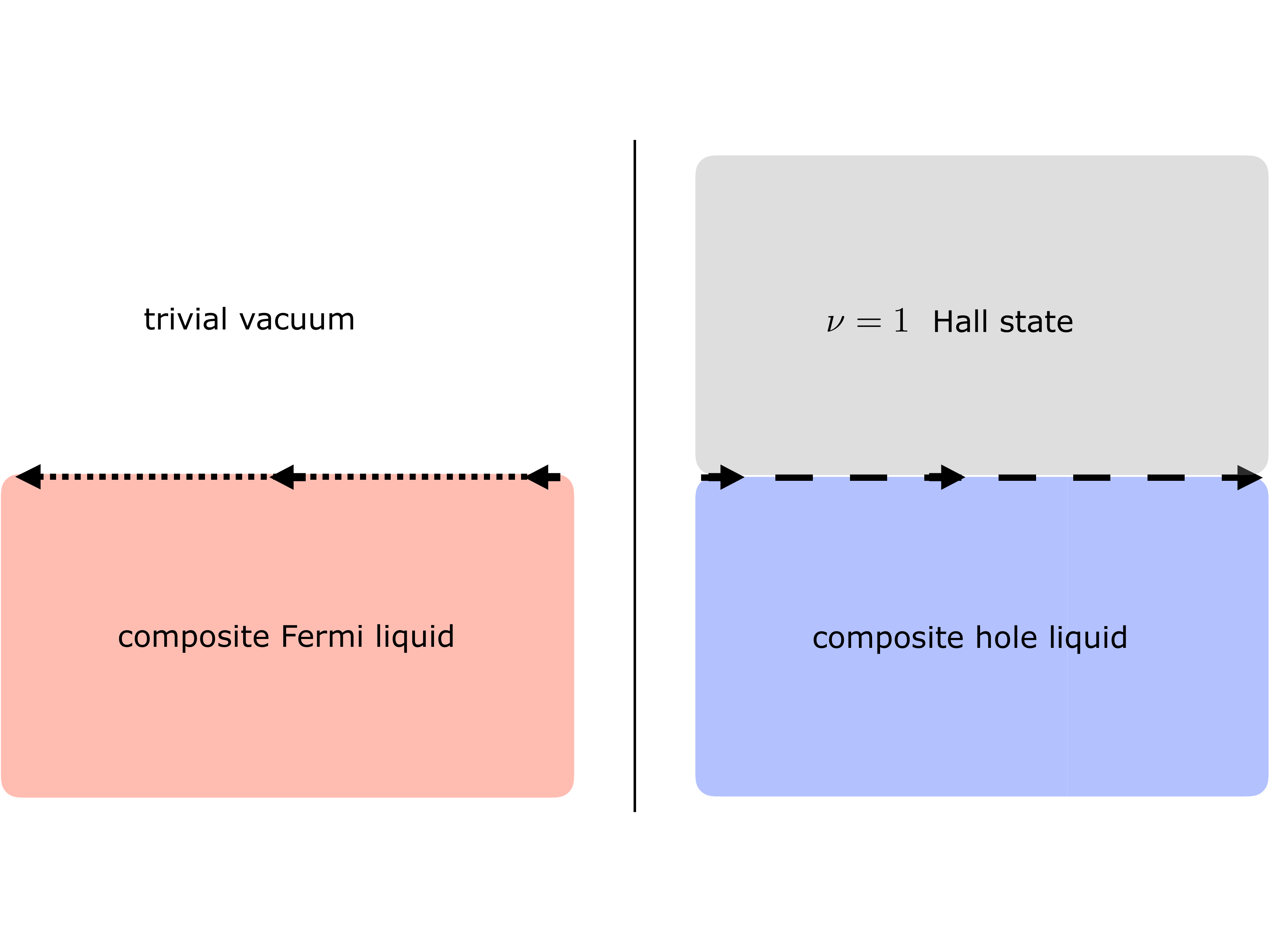}\\
\caption{The left side of the figure depicts the CFL in the lower half-plane and the topologically trivial vacuum in the upper half-plane.
The chiral field $\phi_f$ is indicated by the short dashed line and described by the lagrangian in Eq. (\ref{boundarycompositefermionlagrangian}).
The right side displays the CHL in the lower half-plane and the $\nu=1$ integer quantum Hall state in the upper half-plane.
The chiral field $\phi_h$ is indicated by the long dashed line and described by the lagrangian in Eq. (\ref{boundarycompositeholelagrangian}).
A particle-hole transformation maps the left and right configurations into one another and transforms the part of the lagrangians describing $\phi_f$ and $\phi_h$ into one another.}
\label{cfllowtrivupptrans}
\end{figure}

In order to find a consistent definition for how a PH transformation should act on the boundary degrees of freedom, it is useful to consider a setup where the CFL lies in the lower half-plane and the topologically trivial vacuum is in the upper-half plane (Fig. \ref{cfllowtrivupptrans}).
A particle-hole transformation maps this system onto a filled Landau level for $y>0$ and a CHL in $y<0$.
Thus, we require
\begin{align}
\label{zbounddef}
 \phi_f  & \mapsto \phi_h - (1 - \chi ) {\pi \over 4}, \cr
 \phi_h  & \mapsto \phi_f  - (1 + \chi) {\pi \over 4}, \cr
\phi_{\rm LL}  & \mapsto - \phi_{\rm LL} + \chi {\pi \over 2},
\end{align}
where $\chi = +1$ for a right-moving field and $\chi = -1$ for a left-moving field.
(The choice for the constant shifts of the fields is not determined from consistency of the above picture, but rather from a consistent transformation rule for the emergent composite fermion described in the next section.)
At first glance, the resulting description seems to miss the mode corresponding to the filled Landau level. 
However, as shown in Fig. \ref{cfllowtrivupptrans}, the filled Landau level occurs over the entire 2D plane. 
Thus, generally, it thus does not contribute any new edge mode.

\begin{figure}[h!]
\centering
\includegraphics[width=.7\linewidth]{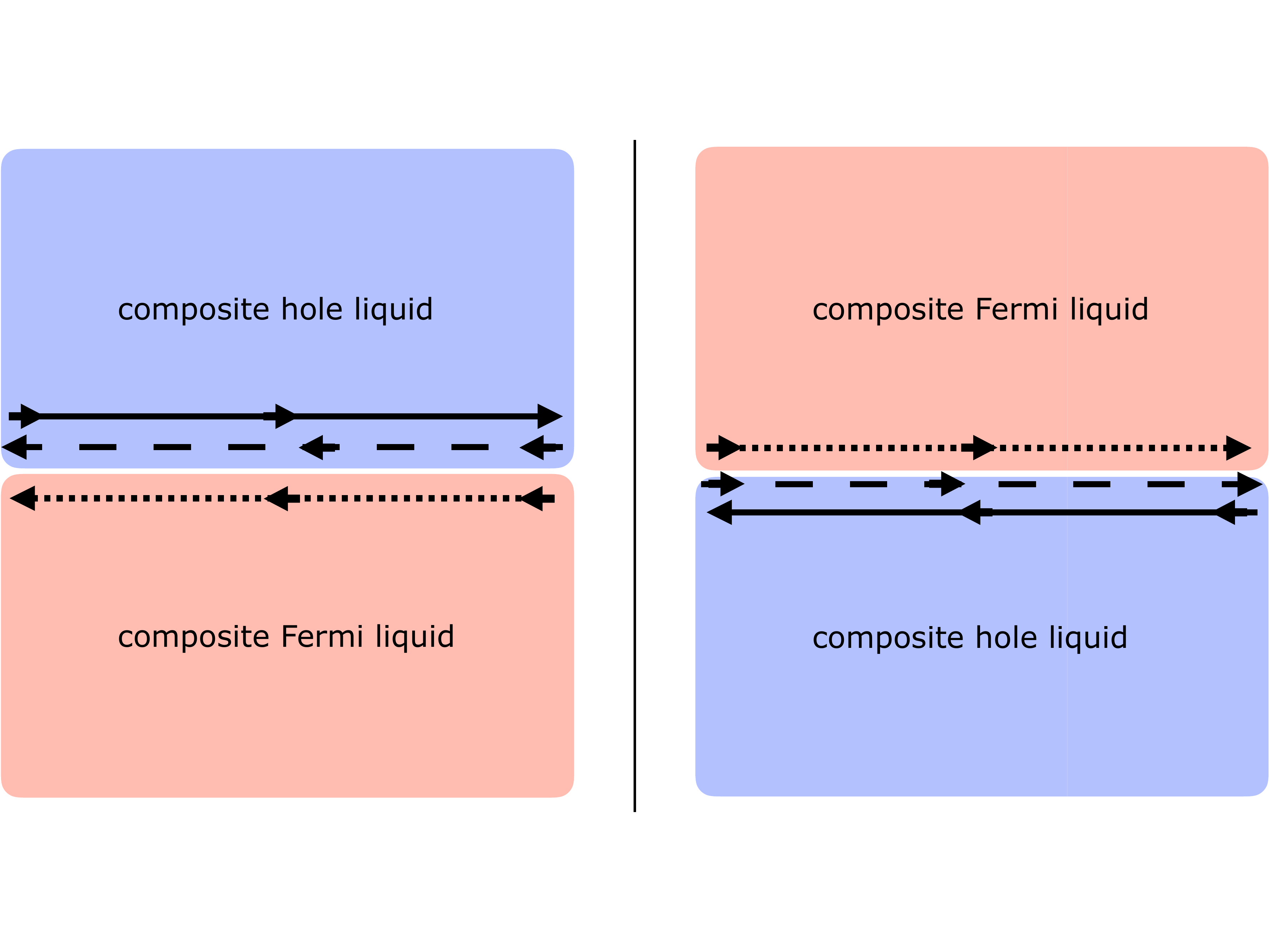}\\
\caption{The left side of the figure depicts the CFL in the lower half-plane and the CHL in the upper half-plane.
The right side displays the CFL in the lower half-plane and the CHL in the upper half-plane.
The chiral field $\phi_f$ is indicated by the short dashed line, the $\phi_h$ field by the long dashed line, and the chiral mode $\phi_{\rm LL}$ of the filled Landau level by the full line.
A particle-hole transformation maps the left and right configurations into one another.}
\label{cfllowchlup}
\end{figure}

We infer the PH transformation on the filled Landau level mode using a setup where the composite Fermi liquid is in the lower-half plane and the composite hole liquid is in the upper half-plane (Fig. \ref{cfllowchlup}).
Under a PH transformation, the chirality of the Landau level mode is reversed, but the sign of the coupling to electromagnetism is maintained.
This ensures that the electromagnetic charge of $\psi_{\rm LL}$ is preserved since an initially right-moving Landau level electron creation operator along the interface $\psi_{\rm LL}^\dagger \sim e^{i \phi_{\rm LL}}$ maps to a left-moving destruction operator $\psi_{\rm LL} \sim e^{i \phi_{\rm LL}}$ under a PH transformation in this particular setup.

\section{Emergence of a particle-hole symmetric effective theory}
\label{sec:p-hsymemergence}
\subsection{General setup}

\begin{figure}[h!]
\centering
\includegraphics[width=.7\linewidth]{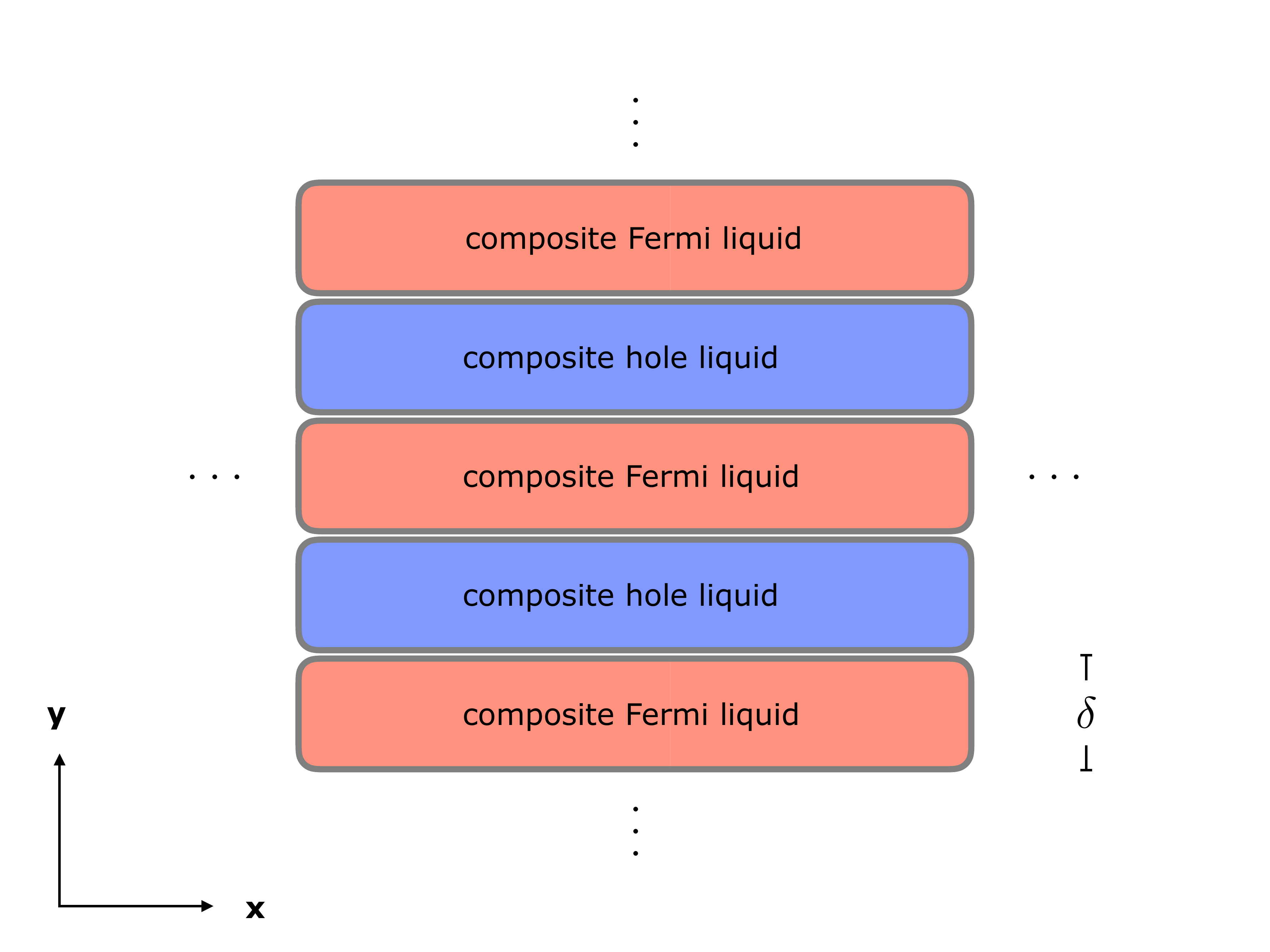}\\
\caption{Schematic picture of the two-dimensional spatial plane patterned with alternating strips of the composite Fermi liquid -- shown in red -- and the composite hole liquid -- shown in blue.
The strips have width equal to $\delta$ and run along the $x$-direction.}
\label{patternedplane}
\end{figure}

Having reviewed the description of a half-filled Landau level in terms of composite electrons or composite holes, we now describe how a PH symmetric state arises from alternating strips of the composite Fermi and composite hole liquid.
The specific setup that we consider is illustrated in Fig. \ref{patternedplane}.
It consists in taking the CFL to lie in narrow strips of width $\delta$ oriented along the $x$-direction for $y \in [q \delta, (q+1)\delta]$ with $q \in 2 \mathbb{Z} + 1$ and the CHL to lie in the strips between $y \in [p \delta, (p + 1)\delta]$ with $p \in 2 \mathbb{Z}$.
This setup is invariant under translations along the $x$-direction and translations by $2 \delta$ along the $y$-direction, but it is not invariant under general spatial rotations.

This configuration is not invariant under a PH transformation. 
However, it is invariant under the combined action of a PH transformation and a translation by one unit equal to $\delta$ in the $y$-direction.
We refer to this combined action as the {\it{microscopic particle-hole transformation}}.
At long wavelengths, we will see that the combined action becomes identical to a PH transformation.

\subsection{Coupled-wire approach}

The long wavelength theory is obtained in the limit where the strip width $\delta \rightarrow 0$.
In this limit, the physics of the CFL and CHL regions effectively becomes one-dimensional.
Thus, it is natural to describe the composite fermions and composite holes within their respective narrow strips in terms of wire arrays. 

We consider a setup in which there are $N_f$ ($N_h$) composite Fermi (hole) wires within each composite Fermi (hole) strip.
The wires are taken to run along the $x$-direction and to lie at regularly spaced intervals between $y/\delta \in \mathbb{Z}$.
On each wire, indexed by $y$, it is convenient to expand $f^\dagger$ and $h^\dagger$ in terms of their chiral components; for convenience, we also write out the form of the electron creation operator of the filled Landau level mode:
\begin{align}
\label{lineexpansion}
f^\dagger(t,x,y) & = e^{- i k^{L}_f (y) x} f_L^\dagger(t,x,y) + e^{- i k^{R}_f (y) x} f_R^\dagger(t,x,y), \cr
h^\dagger(t,x,y) & = e^{- i k^{L}_h(y) x} h_L^\dagger(t,x,y) + e^{- i k^{R}_h(y) x} h_R^\dagger(t,x,y), \cr
\psi_{\rm LL}^\dagger(t,x,y) & = \gamma_y e^{-i k_{{\rm LL}}(y) x} e^{(-1)^{{y \over \delta}} i \phi_{\rm LL}(t,x,y)}.
\end{align}
The Klein factors ($\gamma_y = \gamma_y^\ast$) obey the anti-commutation relations: $\{\gamma_y, \gamma_{y'} \} = 2 \delta_{y,y'}, \{\gamma_y, f \} = 0$, and $\{\gamma_y, h \} = 0$.
The differing factors of $e^{i k_{f,h} x}$ in the above expansions are due to the fact that $f$ and $h$ couple to electromagnetism with opposite sign and the possibility that $N_f \neq N_h$.

On each wire, the 1D composite fermion density -- identified with the 1D limit of the electron fluid density -- equals ${k_f^{R}(y) - k_f^{L}(y) \over 2 \pi}$ (and similarly for the quasi-hole density on each wire).
The half-filling condition on the electrons or holes combined with the existence of $N_f$ $(N_h)$ wires per length $\delta$ in the $y$-direction implies
\begin{align}
\label{compositefermi}
k_f^{L/R}(y) & =  \mp {|B| \delta \over 4 N_f} - B y, \cr
k_h^{L/R}(y) & = \mp {|B| \delta \over 4 N_h} + B y.
\end{align}
The position-dependent shift proportional to $By$ of the Fermi points away from their zero field values at $\pm (k_f^{R}(y) - k_f^{L}(y))/2$ (and similarly for the holes) implies that we do not assume the emergent gauge fields cancel the background magnetic flux.
In other words, we do not assume that the composite Fermi or composite hole liquids realize the mean-field solution to the equations of motion in which the emergent gauge field screens the background magnetic flux.
Away from half-filling, the one-dimensional composite Fermi and composite hole densities generally differ.

To describe the filled Landau level, we take $N_{\rm LL} \geq 2$ free fermion wires regularly spaced at $y/\delta \in 2 \mathbb{Z} + {n \over N_{\rm LL} - 1}$ for $n = 0, 1, \ldots, N_{\rm LL} - 1$ within each composite hole region.\footnote{The assumption of non-interacting fermions is inessential and is merely made to simplify the description. Weak interactions may be included using bosonization.}
Within these regions, the wires are coupled together via appropriate inter-wire integer Hall state generating perturbations.\footnote{In the linearized limit, such perturbations are identified as the unique single-particle interaction that scatters a left-moving fermion into a right-moving fermion on the wire above it while maintaining translations along the $x$-direction and charge conservation.}
There are no free fermion wires in the composite Fermi regions.
The Fermi momenta of the filled Landau level edge modes:
\begin{align}
\label{filledLLfermiwaveeven}
k_{\rm LL}(y) = {|B| \delta \over 2 (N_{\rm LL} - 1)} - B y, 
\end{align}
for the right-movers at $y/\delta \in 2 \mathbb{Z}$, while
\begin{align}
\label{filledLLfermiwaveodd}
k_{\rm LL}(y) = - {|B| \delta \over 2 (N_{\rm LL} - 1)} - B y, 
\end{align}
for the left-movers at $y/\delta \in 2 \mathbb{Z} + 1$.

While the numbers of wires $N_f, N_h$ and $N_{\rm LL}$ within each region are free parameters, we find it necessary to make a specific choice in the analysis presented below.
We take $N_f = N_h = (N_{\rm LL} - 1)/2 = 1.$
From the above equations, we see that this choice implies the 1D densities of the modes on each wire are identical; $N_f = N_h = 1$ is simply the minimal number of wires required and naturally obtains in the one-dimensional $\delta \rightarrow 0$ limit of interest.
We will show that this choice ensures that the interactions we study preserve translation invariance along the wires; for general values of $N_f, N_h$ and $N_{\rm LL}$, our route to a PH symmetric composite Fermi liquid requires translation-invariance to be broken along the wires.\footnote{It is generally an interesting, unsolved question to determine the degree to which disorder is important to the stabilization of the metallic behavior observed about half-filing in the two-dimensional electron gas.
We do not consider the role played by weak (or strong, for that matter) disorder in this paper, however, such considerations are well worth further study.}

The emergent gauge fields are not restricted to the wires; they can be taken to live throughout each narrow strip.
Thus, $a_\mu$ lives in the composite Fermi strips $x \in (-\infty, \infty), y \in [q \delta, (q+1) \delta]$ for $q \in 2 \mathbb{Z} + 1$, while $b_\mu$ is defined in the composite hole regions $x \in (-\infty, \infty), y \in [p \delta, (p+1) \delta]$ for $p \in 2 \mathbb{Z}$.

\subsubsection{Coupled-wire effective action}

\begin{figure}[h!]
\centering
\includegraphics[width=.7\linewidth]{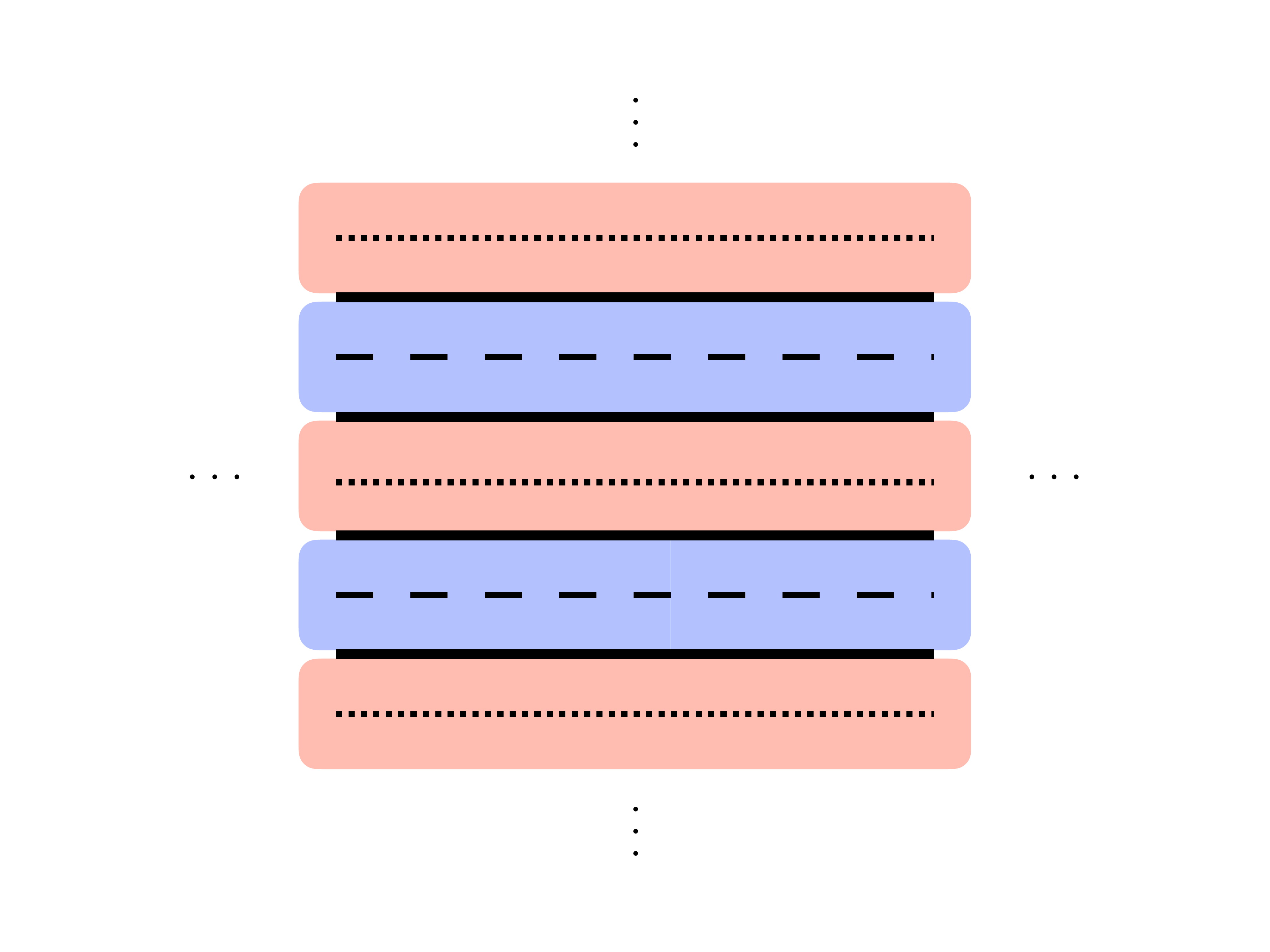}\\
\caption{Wires hosting composite fermions are depicted by short-dashed lines along $y/\delta \in 2 \mathbb{Z} - {1 \over 2}$ within the (red) composite Fermi liquid regions, while wires hosting composite holes are indicated by long-dashed lines along $y/\delta \in 2 \mathbb{Z} + {1 \over 2}$ within the (blue) composite hole liquid regions.
The solid lines along $y/\delta \in \mathbb{Z}$ denote line interfaces between the composite Fermi and composite hole liquids with excitations described by the action $S_{\rm interfaces}$.
The solid lines host the co-propagting left-moving (right-moving) fields $\phi_f$ and $\phi_h$ and the counter-propagating right-moving (left-moving) filled Landau level mode $\phi_{\rm LL}$ for $y/\delta \in 2 \mathbb{Z}$ ($y/\delta \in 2 \mathbb{Z} + 1$).}
\label{coupledwires}
\end{figure}

Writing out the low-energy effective action $S = S_{\rm bulk} + S_{\rm interfaces}$ for the patterned plane in terms of the emergent gauge fields and one-dimensional wire degrees of freedom introduced above, we find
\begin{align}
\label{fullbulk}
S_{\rm bulk} & = \sum_{y/\delta \in 2\mathbb{Z} - {1 \over 2}} \int_{t,x} \Big[f^\dagger_{L} i \Big(D^f_t - v_f D^f_x\Big) f_L + f^\dagger_{R} i \Big(D^f_t + v_f D^f_x\Big) f_R \Big] \cr
& + {1 \over 4 \pi} \sum_{q \in 2 \mathbb{Z} + 1} \int_{t,x,} \int_{q \delta}^{(q+1) \delta} dy \Big[{1 \over 2} \epsilon_{\mu \nu \rho} a_\mu \partial_\nu a_\rho \Big] \cr
& + \sum_{y/\delta \in 2 \mathbb{Z} + {1 \over 2}} \int_{t,x} \Big[ h^\dagger_{L} i \Big(D^h_t - v_h D^h_x\Big) h_L + h^\dagger_{R} i \Big(D^h_t + v_h D^h_x\Big) h_R \Big] \cr
& + {1 \over 4 \pi} \sum_{p \in 2 \mathbb{Z}} \int_{t,x,} \int_{p \delta}^{(p+1) \delta} dy \Big[- {1 \over 2} \epsilon_{\mu \nu \rho} b_\mu \partial_\nu b_\rho + \epsilon_{\mu \nu \rho} A_\mu \partial_\nu A_\rho \Big] 
\end{align}
and
\begin{align}
\label{fullinterfaces}
S_{\rm interface} & = {1 \over 4 \pi} \sum_{y/\delta \in \mathbb{Z}} \int_{t, x} \Big[2 (- 1)^{{y \over \delta}} \Big(\partial_t \phi_f +{(-1)^{{y \over \delta}} \over 2} a_t\Big) \Big(\partial_x \phi_f + {(-1)^{{y \over \delta}} \over 2} a_x\Big) - v_{\phi_f} \Big(\partial_x \phi_f + {(-1)^{{y \over \delta}} \over 2} a_x\Big)^2  \cr 
+ & \epsilon_{\mu \nu y} a_\mu \partial_\nu \phi_f + 2 (-1)^{{y \over \delta}} \Big(\partial_t \phi_h + {(-1)^{{y \over \delta}} \over 2} b_t\Big) \Big(\partial_x \phi_h + {(-1)^{{y \over \delta}} \over 2} b_x\Big) - v_{\phi_h} \Big(\partial_x \phi_h + {(-1)^{{y \over \delta}} \over 2} b_x\Big)^2 \cr 
& + \epsilon_{\mu \nu y} b_\mu \partial_\nu \phi_h + \partial_x \phi_{\rm LL} \Big((-1)^{{y + \delta \over \delta}} \partial_t - v_{\rm LL} \partial_x\Big) \phi_{\rm LL} + 2 \epsilon_{\mu \nu y} A_\mu \partial_\nu \phi_{\rm LL} \Big].
\end{align}
Fig. \ref{coupledwires} depicts more precisely our specific setup.
We have taken the CFL and CHL wires to lie at $y/\delta \in 2 \mathbb{Z} \mp {1 \over 2}$.
Since $N_f = N_h = 1$, there is no inter-wire composite fermion tunneling within a CFL or CHL region.
The covariant derivatives: ${\bf{D}}^f \equiv {\mathbf{\partial}} - i ({\bf{a}} + {\bf{A}})$ and ${\bf{D}}^h \equiv {\bf{\partial}} - i ({\bf{b}} - {\bf{A}})$.
The Fermi velocities: $v_f = k_F/m_f$ and $v_h = k_F/m_h$.
Invariance under the microscopic PH transformation requires all velocities to be independent of position and to satisfy: $v_f = v_h \equiv v_F$ and $v_{\phi_f} = v_{\phi_h} \equiv v$.

To study the action $S$, it is convenient to make the following field redefinitions along the line interfaces located at $y/\delta \in \mathbb{Z}$,
\begin{align}
\label{fieldredefs}
\phi_f & = {\varphi + \tilde{\varphi} \over 2}, \cr
\phi_h & = {\varphi - \tilde{\varphi} \over 2}, \cr
a_\mu & = \alpha_\mu - \beta_\mu, \cr
b_\mu & = \alpha_\mu + \beta_\mu,
\end{align}
in terms of which action along the interfaces becomes
\begin{align}
S_{\rm interfaces} & = {1 \over 4 \pi} \sum_{y/\delta \in \mathbb{Z}} \int_{t, x} \Big[(- 1)^{{y \over \delta}} \Big(\partial_t \varphi + (-1)^{{y \over \delta}} \alpha_t\Big) \Big(\partial_x \varphi + (-1)^{{y \over \delta}} \alpha_x\Big) - {v \over 2} \Big(\partial_x \varphi + (-1)^{{y \over \delta}} \alpha_x\Big)^2  \cr 
+ & \epsilon_{\mu \nu y} \alpha_\mu \partial_\nu \varphi + (-1)^{{y \over \delta}} \Big(\partial_t \tilde{\varphi} + (-1)^{{y + \delta \over \delta}} \beta_t\Big) \Big(\partial_x \tilde{\varphi} + (-1)^{{y + \delta \over \delta}} \beta_x\Big) - {v \over 2} \Big(\partial_x \tilde{\varphi} + (-1)^{{y + \delta \over \delta}} \beta_x\Big)^2 \cr 
& - \epsilon_{\mu \nu y} \beta_\mu \partial_\nu \tilde{\varphi} + \partial_x \phi_{\rm LL} \Big((-1)^{{y + \delta \over \delta}} \partial_t - v_{\rm LL} \partial_x\Big) \phi_{\rm LL} + 2 \epsilon_{\mu \nu y} A_\mu \partial_\nu \phi_{\rm LL} \Big].
\end{align}
Note that $\varphi$ carries $+1$ ($-1$) charge -- it transforms as $\varphi \mapsto \varphi + (-1)^{{y + \delta \over \delta}} {\Lambda_a + \Lambda_b \over 2}$ -- under an $\alpha_\mu$ gauge transformation, while $\tilde{\varphi}$ carries $-1$ ($+1$) charge -- it transforms as $\tilde{\varphi} \mapsto \tilde{\varphi} + (-1)^{{y \over \delta}} {\Lambda_b - \Lambda_a \over 2}$ -- under a $\beta_\mu$ gauge transformation for $y/\delta \in 2 \mathbb{Z} + 1$ ($y/\delta \in 2 \mathbb{Z}$).
In the next section, we will argue that $\varphi$ describes the low-energy excitations of a PH symmetric composite Fermi liquid.

Within our coupled-wire setup, the electron, hole, and filled Landau level boundary or interface creation operators defined at $y/\delta \in \mathbb{Z}$ take the form,
\begin{align}
\label{boundopform}
\psi_e^\dagger(t,x,y) & = e^{(-1)^{{y + \delta \over \delta}} i (\varphi + \tilde{\varphi})(y)} f^\dagger(y \mp {\delta \over 2}), \cr
\psi_h^\dagger(t,x,y) & = e^{(-1)^{{y + \delta \over \delta}} i (\varphi - \tilde{\varphi})(y)} h^\dagger(y \pm {\delta \over 2}), \cr
\psi_{\rm LL}^\dagger(t,x,y) & = \gamma_y e^{- i k_{{\rm LL}}(y) x} e^{(-1)^{{y \over \delta}} i \phi_{\rm LL}(y)},
\end{align}
for $y/\delta \in 2 \mathbb{Z}$ (upper) or $y/\delta \in 2 \mathbb{Z} + 1$ (lower).
$\psi_e^\dagger$ may be said to create $\varphi$ and $\tilde{\varphi}$ excitations along with the composite fermion.
In contrast, $\psi_h^\dagger$ creates a $\varphi$ excitation, annihilates a $\tilde{\varphi}$ excitation, and creates a composite hole.
Note that we are working in the $\alpha_y = \beta_y =0$ gauge.

Using Eq. (\ref{zbounddef}), the fields introduced in Eq. (\ref{fieldredefs}) transform as follows under the PH transformation:
\begin{align}
\label{znewfields}
\varphi & \mapsto \varphi - {\pi \over 2}, \cr
\tilde{\varphi} & \mapsto - \tilde{\varphi} + \chi {\pi \over 2}, \cr
(\alpha_t, \alpha_x, \alpha_y) & \mapsto (\alpha_t, - \alpha_x, - \alpha_y), \cr
(\beta_t, \beta_x, \beta_y) & \mapsto (- \beta_t, \beta_x, \beta_y).
\end{align}
Recall that $\chi = \pm 1$ for a right/left-moving field.
Thus, $\tilde{\varphi}$ transforms identically to the filled Landau level mode $\phi_{\rm LL}$, $\alpha_\mu$ transforms by time-reversal, and $\beta_\mu$ transforms by the combination of charge-conjugation and time-reversal, just like the electromagnetic field.
The microscopic PH transformation supplements the above action with a translation of the $y$-coordinate of the fields by $\delta$.
Using Eqs. (\ref{boundopform}) and (\ref{znewfields}), the microscopic PH transformation takes
\begin{align}
\label{boundoptrans}
\psi_e^\dagger(t,x,y) & \mapsto (-1)^{{y + \delta \over \delta}} \psi_h^\dagger(t,x,y+\delta), \cr
\psi_h^\dagger(t,x,y) & \mapsto (-1)^{{y \over \delta}} \psi_e^\dagger(t,x,y+\delta), \cr
\psi_{\rm LL}^\dagger(t,x,y) & \mapsto - i \psi_{\rm LL}(t,x,y+\delta).
\end{align}

\subsection{Emergence of the particle-hole symmetric composite fermion}

We now turn to a description for how a PH symmetric composite Fermi liquid can emerge from $S$, supplemented by suitable interactions.
In total, the symmetries we wish to preserve are (i) electromagnetism, (ii) translations along the $x$-direction, (iii) translations by $2 \delta$ in the $y$-direction, and (iv) invariance under the microscopic PH transformation.

\subsubsection{The interaction}

In the spirit of effective field theory, we generally allow all interactions consistent with the symmetries of $S$.
After some exploration, we have found it useful to concentrate upon the effects of the following interaction $H_{\rm int} = H_1 + H_2$:
\begin{align}
H_{1} & = g_1 \sum_{y/\delta \in \mathbb{Z}} (-1)^{{y \over \delta}} \Big(\psi_e^\dagger \psi_h \psi_{\rm LL} \partial_x \psi_{\rm LL} \Big) (y) + {\rm h.c.} \cr
H_2 & = g_2 \sum_{y/\delta \in 2 \mathbb{Z}} \Big(\psi_e^\dagger \psi_{\rm LL} \Big)(y + \delta) \Big( \psi_h \psi_{\rm LL} \Big) (y) - g_2 \sum_{y/\delta \in 2 \mathbb{Z} + 1} \Big(\psi_h^\dagger \psi^\dagger_{\rm LL} \Big)(y + \delta) \Big( \psi_e \psi^\dagger_{\rm LL} \Big) (y)  + {\rm h.c.} \cr
\end{align}
We will show that study of $H_{\rm int}$ provides a saddle-point from which a PH symmetric composite Fermi liquid emerges.
The couplings $g_1$ and $g_2$ are real.
We verify using Eq. (\ref{boundoptrans}) that $H_{\rm int}$ is invariant under the microscopic PH transformation.
Using the expansions and definitions in Eqs. (\ref{lineexpansion}) - (\ref{filledLLfermiwaveodd}), the interaction reduces to
\begin{align}
\label{g1operator}
H_1 & = g_1 \sum_{y/\delta \in 2 \mathbb{Z}} e^{- 2 i \Big(\tilde{\varphi}(y) + \phi_{\rm LL}(y) \Big)} \Big[f_L^\dagger(y - {\delta \over 2}) h_R(y + {\delta \over 2}) + f^\dagger_R(y - {\delta \over 2}) h_L(y + {\delta \over 2}) \Big] \cr
& - g_1 \sum_{y/\delta \in 2 \mathbb{Z} + 1} e^{2 i \Big(\tilde{\varphi}(y) + \phi_{\rm LL}(y) \Big)} \Big[f_L^\dagger(y + {\delta \over 2}) h_R(y - {\delta \over 2}) + f^\dagger_R(y + {\delta \over 2}) h_L(y - {\delta \over 2}) \Big],
\end{align}
\begin{align}
H_2 & = g_2 \sum_{y/\delta \in 2 \mathbb{Z}} e^{i \Big(\varphi(y+\delta) + \tilde{\varphi}(y+ \delta) + \phi_{\rm LL}(y+ \delta) \Big)} e^{i \Big(\varphi(y) - \tilde{\varphi}(y) - \phi_{\rm LL}(y) \Big)} \gamma_{y+\delta} \gamma_y \cr
& \times \Big[f_L^\dagger (y + {3 \delta \over 2}) h_L(y + {\delta \over 2}) + f_R^\dagger (y+{3 \delta \over 2}) h_R(y + {\delta \over 2}) \Big] \cr
& - g_2 \sum_{y/\delta \in 2 \mathbb{Z} + 1} e^{- i \Big(\varphi(y+\delta) - \tilde{\varphi}(y+ \delta) - \phi_{\rm LL}(y+ \delta) \Big)} e^{- i \Big(\varphi(y) + \tilde{\varphi}(y) + \phi_{\rm LL}(y) \Big)} \gamma_{y+\delta} \gamma_y \cr
& \times \Big[h_L^\dagger (y + {3 \delta \over 2}) f_L(y + {\delta \over 2}) + h_R^\dagger (y+{3 \delta \over 2}) f_R(y + {\delta \over 2}) \Big]  + {\rm h.c.}
\end{align}
in the long wavelength effective theory.
In the above expansion, we have only retained terms without oscillating prefactors.
Note that we have had to assume that the composite electrons and holes are at half-filling with respect to an underlying spatial lattice running along the $x$-direction in order to obtain these expansions.

We imagine working in the perturbative limit $g_2 \ll g_1 \ll 1$ which allows us to first focus upon the effects of $H_1$ and to subsequently study the effects of $H_2$.
This limit may be justified by noting that $H_1$ is an ``on-site" interaction for fields localized near a line interface separating a composite Fermi and composite hole region, while $H_2$ connects fields across a strip.
We show below that $H_1$ patches together the composite Fermi and composite hole strips.
$H_2$ both delocalizes the electrically-neutral composite fermion arising from the fermionization of $\varphi$ and lifts the composite fermions $f$ and $h$ from the low-energy effective theory.

\subsubsection{Patching the strips together}

We first consider the effects of $H_1$.
About the non-interacting point in parameter space, $H_1$ has scaling dimension equal to two and is marginal. 
We assume the presence of short-ranged repulsive density-density interactions between $\tilde{\varphi}$ and $\phi_{\rm LL}$ which drive $H_1$ relevant.
Formally integrating out the composite fermions $f$ and $h$ at one-loop, we generate a (Euclidean signature) effective potential for the operator ${\cal O}_1(t,x,y) = e^{2 i \Big(\tilde{\varphi}(t, x,y) + \phi_{\rm LL}(t,x,y) \Big)}$,
\begin{align}
V_{{\cal O}_1} = g_1^2 \int dk_0 dk_x dy\ \log({\sqrt{k_0^2 + k_x^2} \over \Lambda}) {\cal O}_1^\ast(-k_0, - k_x, y) {\cal O}_1(k_0, k_x, y).
\end{align}
In writing the above potential, we have performed a partial Fourier transform with respect to the Euclidean time and $x$-coordinate.
The cutoff $\Lambda$ can be identified with the inverse of a short-distance cutoff, such as an underlying spatial lattice.

The one-loop potential $V_{{\cal O}_1}$ has an instability in the infrared ${\sqrt{k_0^2 + k_x^2} \over \Lambda} \ll 1$ which favors the condensation  $\langle {\cal O}_1 \rangle \neq 0$.
Invariance under the microscopic PH transformation implies that the condensation occurs by fixing
\begin{align}
\label{condensate}
\langle \tilde{\varphi} + \phi_{\rm LL} \rangle= 0.
\end{align}
Thus, we infer from the form of the one-loop potential that the IR dynamics are such that Eq. (\ref{condensate}) is satisfied.

What are the consequences of the condensate $\langle  \tilde{\varphi} + \phi_{\rm LL} \rangle = 0$?
Under the emergent and electromagnetic gauge symmetries, ${\cal O}_1$ transforms as
\begin{align}
{\cal O}_1(t,x,y) \mapsto e^{(-1)^{{y + \delta \over \delta}} i (\Lambda_a + \Lambda_A) - i (\Lambda_b - \Lambda_A)} {\cal O}_1(t,x,y)
\end{align}
where $\Lambda_a$ ($\Lambda_b$) is the gauge parameter of $a_\mu$ ($b_\mu$) and $\Lambda_A$ is the parameter for electromagnetism.
Therefore, if ${\cal O}_1$ (or more precisely, $|{\cal O}_1|^2$) acquires a vacuum expectation value, then the expansion about the resulting minimum of the effective potential results in a breaking of the full gauge symmetry to the subgroup under which ${\cal O}_1$ is neutral.
Thus, we equate
\begin{align}
\label{patching}
a_\mu + A_\mu = b_\mu - A_\mu 
\end{align}
along the line interfaces.
Eq. (\ref{patching}) is the patching condition that defines a single emergent gauge field.
This equation can equivalently be written as
\begin{align}
\beta_\mu = A_\mu
\end{align}
and allows us to identify $\alpha_\mu$ as the globally-defined emergent gauge field.
(The electromagnetic field is, of course, already defined throughout the 2D plane.)
We can check using Eq. (\ref{znewfields}) that the patching condition respects the PH transformation.
The second consequence is that the $\tilde{\varphi}$ and $\phi_{\rm LL}$ excitations obtain a mass and decouple from the low-energy physics. 
The third implication is simply that when we replace ${\cal O}_1$ by its vacuum expectation value, i.e., set $\tilde{\varphi} + \phi_{\rm LL} = 0$ in Eq. (\ref{g1operator}), the composite fermions $f$ and $h$ can hop into one another across the 2D plane.

From the above considerations, $S = S_{\rm bulk} + S_{\rm interfaces}$ simplifies to
\begin{align}
\label{fullbulksimplified}
S_{\rm bulk} 
& = {1 \over 4 \pi} \sum_{m \in \mathbb{Z}} \int_{t,x} \int_{m \delta}^{(m+1) \delta} dy \Big[{(-1)^{m+1} \over 2} \epsilon_{\mu \nu \rho} \alpha_\mu \partial_\nu \alpha_\rho - \epsilon_{\mu \nu \rho} A_\mu \partial_\nu \alpha_\rho + {1 \over 2} \epsilon_{\mu \nu \rho} A_\mu \partial_\nu A_\rho \Big] \cr
& + \sum_{y/\delta \in 2\mathbb{Z} - {1 \over 2}} \int_{t,x} \Big[f^\dagger_{L} i \Big(D_t - v_F D_x\Big) f_L + f^\dagger_{R} i \Big(D_t + v_F D_x\Big) f_R \Big] \cr
& + \sum_{y/\delta \in 2 \mathbb{Z} + {1 \over 2}} \int_{t,x} \Big[ h^\dagger_{L} i \Big(D_t - v_F D_x\Big) h_L + h^\dagger_{R} i \Big(D_t + v_F D_x\Big) h_R \Big] 
\end{align}
for $D_\mu \equiv \partial_\mu - i \alpha_\mu$ and
\begin{align}
S_{\rm interfaces} & = {1 \over 4 \pi} \sum_{y/\delta \in \mathbb{Z}} \int_{t, x} \Big[(- 1)^{{y \over \delta}} \Big(\partial_t \varphi + (-1)^{{y \over \delta}} \alpha_t\Big) \Big(\partial_x \varphi + (-1)^{{y \over \delta}} \alpha_x\Big) \cr 
& - {v \over 2} \Big(\partial_x \varphi + (-1)^{{y \over \delta}} \alpha_x\Big)^2  + \epsilon_{\mu \nu y} \alpha_\mu \partial_\nu \varphi  \Big].
\end{align}
In $S_{\rm bulk}$, we note the appearance of a Chern-Simons term for $\alpha_\mu$ with alternating coefficient proportional to $(-1)^{m+1}$, a $BF$ coupling between $\alpha_\mu$ and the electromagnetic field $A_\mu$, and a level $1/2$ Chern-Simons term for $A_\mu$.
In the long wavelength limit $\delta \rightarrow 0$, the alternating Chern-Simons term can be ignored.
Therefore, we will henceforth drop this term in the analysis below.
$S_{\rm interfaces}$ describes a collection of one-dimensional wires each hosting a chiral boson minimally coupled to $\alpha_\mu$ whose chirality alternates from wire to wire.

\subsubsection{Delocalization of the neutral composite fermion}

To study $H_2$ evaluated about the saddlepoint $\tilde{\varphi} + \phi_{\rm LL} = 0$,
it is convenient to fermionize the $\varphi$ excitations by defining
\begin{align}
\psi^\dagger_L(y) & \equiv \gamma_y e^{- i \varphi(y)},{\rm for}\ y/\delta \in 2 \mathbb{Z} \cr
\psi^\dagger_R(y) & \equiv \gamma_y e^{i \varphi(y)}, {\rm for}\ y/\delta \in 2 \mathbb{Z} + 1.
\end{align}
in terms of which
\begin{align}
H_2 = & g_2 \sum_{y/\delta \in 2 \mathbb{Z}} \psi_R^\dagger(y+\delta) \psi_L(y) \Big[f_L^\dagger (y + {3 \delta \over 2}) h_L(y + {\delta \over 2}) + \Big(L \leftrightarrow R \Big) \Big] \cr
& - g_2 \sum_{y/\delta \in 2 \mathbb{Z} + 1} \psi_L^\dagger(y+\delta) \psi_R(y) \Big[h_L^\dagger (y + {3 \delta \over 2}) f_L(y + {\delta \over 2}) + \Big(L \leftrightarrow R \Big) \Big]  + {\rm h.c.}
\end{align}
In exact analogy to our analysis of $H_1$, the one-loop effective potential for the chiral operators ${\cal O}_2^{\rm even} \equiv f_L^\dagger (y + {3 \delta \over 2}) h_L(y + {\delta \over 2}) + f_R^\dagger (y+{3 \delta \over 2}) h_R(y + {\delta \over 2})$ for $y/\delta \in 2 \mathbb{Z}$ and ${\cal O}_2^{\rm odd} \equiv h_L^\dagger (y + {3 \delta \over 2}) f_L(y + {\delta \over 2}) + h_R^\dagger (y+{3 \delta \over 2}) f_R(y + {\delta \over 2})$ for $y/\delta \in 2 \mathbb{Z} + 1$ generated by $\psi_{L, R}$ favors the non-zero condensates:
\begin{align}
\label{O2condensates}
\langle {\cal O}_2^{\rm even} \rangle & \neq 0, \cr
\langle {\cal O}_2^{\rm odd} \rangle & \neq 0.
\end{align}
Invariance under the microscopic PH transformation requires $\langle {\cal O}_2^{\rm even} \rangle = \langle {\cal O}_2^{\rm odd} \rangle$.
We normalize this vacuum expectation value to unity.
Diagonalizing the quadratic composite fermion effective lagrangian, including the terms induced by the condensates in Eq. (\ref{O2condensates}), we find that $f$ and $h$ are lifted from the low-energy effective theory.
At the same time, the replacement of the ${\cal O}_2^{\rm even,odd}$ operators by their vacuum expectation values in $H_2$ results in the delocalization or ``liberation" of $\psi_{R,L}$ across the 2D plane.

Suppose we had instead integrated out the composite fermions $f$ and $h$ at one-loop.
Due to $H_1$ in Eq. (\ref{g1operator}) evaluated at $\tilde{\varphi} + \phi_{\rm LL} = 0$, the contribution of the two-dimensional composite fermions to the $\psi_R^\dagger \psi_L$ effective potential is proportional to the (Euclidean) momentum carried by $\psi_R^\dagger \psi_L$ and vanishes in the IR.

From this analysis, we obtain the effective action
\begin{align}
S_{\rm eff} & = \sum_{y/\delta \in 2 \mathbb{Z}} \int_{t,x} \Big[\psi_L^\dagger i (D_t - v D_x) \psi_L \Big] + \sum_{y/\delta \in 2 \mathbb{Z} + 1} \int_{t,x} \Big[\psi_R^\dagger i (D_t + v D_x) \psi_R \Big] \cr
& + g_2 \sum_{y/\delta \in 2 \mathbb{Z}} \int_{t,x} \Big[\psi_R^\dagger(y+\delta) \psi_L (y) - \psi_L^\dagger(y+ 2 \delta) \psi_R(y+\delta) + {\rm h.c.}\Big] \cr
& - {1 \over 4 \pi} \int_{t,x,y} \epsilon_{\mu \nu \rho} A_\mu \partial_\nu \alpha_\rho - {1 \over 2} \epsilon_{\mu \nu \rho} A_\mu \partial_\nu A_\rho \Big].
\end{align}
This is precisely the PH symmetric limit of the composite fermion action conjectured by Son \cite{Son2015} in which the $y$-direction has been discretized.
In the 2D effective theory, the PH transformation acts as time-reversal $\psi_L \mapsto i \psi_R$ and $\psi_R \mapsto - i \psi_L$, consistent with the transformation rule in Eq. (\ref{znewfields}) and the choice of 2D $\gamma$-matrices, $\gamma^t = \sigma_1$, $\gamma^x = i \sigma_2$ and $\gamma^y = i \sigma_3$. 

\subsubsection{The neutral composite fermion as an exciton of semions}

Insertion on a boundary of a semion or single vortex of the composite Fermi liquid is accomplished by the $e^{\mp i \phi_f}$ operator, where the sign in the exponential is determined by the orientation of the boundary.
(Recall the electron creation operator on a boundary takes the form $\psi_e^\dagger = e^{\mp 2 i \phi_f} f^\dagger$.)
The semion operator by itself is not physical because $\phi_f$ is charged under the emergent gauge symmetry and so, strictly speaking, we are restricted to operators formed from gauge-invariant combinations of $\phi_f$ and $f$.
Similarly, in the composite hole liquid, $e^{\pm i \phi_h}$ creates a semion on any boundary of the space.

Along a line interface separating the composite Fermi and composite hole liquids, the operator $e^{\mp i (\phi_f + \phi_h)}$ creates one semion from each of the two theories.
Using the field redefinitions in Eq. (\ref{fieldredefs}), we find
\begin{align}
e^{\mp i (\phi_f + \phi_h)} = e^{\mp i \varphi}.
\end{align}
In this way, we see that the electrically neutral composite fermion $\psi^\dagger_{L/R} \sim e^{\mp i \varphi}$ can be viewed as an exciton of semions of the composite Fermi and composite hole theories.
It is interesting to note the similarity to the description advocated by Wang and Senthil \cite{WangSenthilsecond2016} of the particle-hole symmetric composite fermion as a semion dipole.

\section{Conclusion}
\label{conclusion}

In this paper, we have provided a picture for how a particle-hole symmetric composite fermion theory arises from the composite Fermi and composite hole liquids.
The latter two theories are conjugate to one another under a particle-hole transformation.
The particle-hole symmetric formulation was found by examining the long wavelength description of the two-dimensional spatial plane patterned with alternating strips of the composite Fermi and composite hole liquid.

There are a variety of problems that we plan to consider in the future.

It has recently been argued [\onlinecite{mulliganraghu2015}] that composite fermions play a role in the effective description of the field-tuned superconductor-insulator transition in thin films [\onlinecite{Hebard1990, MasonKapitulnik1999, Kapitulnik2001, Yazdani1995, Breznay2015}].
We plan to use methods similar to those in this paper to formulate a particle-vortex symmetric theory for the transition and the emergent metallic phase.

An alternative possible setup in which to derive a particle-hole symmetric composite fermion theory is to take alternating strips of the Pfaffian [\onlinecite{Moore1991}] and anti-Pfaffian theories [\onlinecite{lee2007antipf},\onlinecite{levin2007}]. 
In this case, the analog of the $H_1$ interaction leaves behind an interface theory with chiral central charge equal to two instead of one, as we found above.
It would be interesting to better understand the resulting long wavelength limit.

The constant subleading correction in the entanglement entropy of a system with abelian topological order can be affected by interactions lying along the entanglement cut [\onlinecite{CanoHughesMulligan2015}].
These interactions take the form of the correlated hopping terms studied in this paper.
It would be interesting to characterize how the entanglement entropy of a gapless system is affected, or rather, determined by such interactions.

Landau level mixing is parameterized by the ratio of the cyclotron frequency to the strength of the average Coulomb interaction.
For realistic (bare) parameters, this ratio is of order one.
Nevertheless, a Hall conductivity $\sigma_{xy} = {1 \over 2} {e^2 \over h}$, consistent with particle-hole symmery [\onlinecite{kivelson1997}], is observed at half-filling [\onlinecite{Shahar1995, Shahar1996, Wong1996}].
On a more theoretical level, the Landau level mixing corrections to the lowest Landau level Hamiltonian were calculated in [\onlinecite{BisharaNayak}] and found to be numerically small for realistic values of parameters.
Could these experiments and theoretical results indicate that the particle-hole symmetric composite Fermi liquid is an attractor of the renormalization group?

In previous work [\onlinecite{BMF2015}], it was argued that the CFL and CHL states are distinct phases of matter analogous to the up and down spin states of an Ising system.  
Since the quantity $h \equiv \left( \nu-1/2 \right)$ acts as a particle-hole symmetry breaking field, its sign determines whether the CFL or CHL is the lower energy ground state.  To the extent that there exists a critical point at which particle-hole symmetry is spontaneously broken at $\nu=1/2$ $(h=0)$, it follows that there can also exist a first-order transition between the CFL and CHL as $h$ changes sign.  However, it was shown in [\onlinecite{SpivakKivelsonemulsions}, \onlinecite{JameiKivelsonSpivak2005}] that such first order transitions are forbidden in the presence of dipolar or Coulomb interactions.  
Instead, a modulated ``stripe" phase consisting of CFL and CHL regions would be expected to have a lower energy.  
Thus, it is conceivable that in this case, the choice of patterning strips of CFL and CHL phases may be more than merely a matter of convenience: it may well be the case that broken translational symmetry of the sort considered here in its simplest form (along the $y$-direction) may well describe the physics of the half-filled Landau level in the presence of long-ranged interactions.  However, when the interactions are short-ranged, it remains quite possible that the stripe phase melts and there could well be a first-order transition between the composite Fermi and composite hole liquids.  
We wish to explore such issues, and also to explore similar aspects of half-filled higher Landau levels in future work.  




\section*{Acknowledgments}

We are grateful to E. Berg, E. Fradkin, C. Kane, S. Kivelson, D. T. Son, and A. Vaezi for helfpul discussions.
We thank E. Fradkin and S. Kivelson for insightful comments on a draft of the paper.
This research was supported in part by the John Templeton Foundation (M.M), the DOE Office of Basic Energy Sciences, contract DE-AC02-76SF00515 (S.R.), the National Science
Foundation under Grant No. DMR-14-04230, and the Caltech
Institute of Quantum Information and Matter, an NSF
Physics Frontiers Center with support of the Gordon and Betty
Moore Foundation (M.P.A.F.).

\bibliography{emergentphbib}{}

\providecommand{\href}[2]{#2}\begingroup\raggedright\begin{thebibliography}{10}

\bibitem{jainCF}
J.~K.Jain, {\em Composite Fermions}.
\newblock Cambridge University Press, 2007.

\bibitem{Jiang1989transportanomalies}
H.~W. Jiang, H.~L. Stormer, D.~C. Isui, L.~N. Pfeiffer, and K.~W. West,
  ``Transport anomalies in the lowest landau level of two-dimensional electrons
  at half-filling,'' \href{http://dx.doi.org/10.1103/PhysRevB.40.12013}{{\em
  Phys. Rev. B} {\bfseries 40} (Dec, 1989) 12013--12016}.
  \url{http://link.aps.org/doi/10.1103/PhysRevB.40.12013}.

\bibitem{willett1990}
R.~L. Willett, M.~A. Paalanen, R.~R. Ruel, K.~W. West, L.~N. Pfeiffer, and
  D.~J. Bishop, ``{Anomalous sound propagation at $\nu=1/2$ in a 2D electron
  gas: Observation of a spontaneously broken translational symmetry?},''
  \href{http://dx.doi.org/10.1103/PhysRevLett.65.112}{{\em Phys. Rev. Lett.}
  {\bfseries 65} (Jul, 1990) 112--115}.

\bibitem{willett1993}
R.~L. Willett, R.~R. Ruel, M.~A. Paalanen, K.~W. West, and L.~N. Pfeiffer,
  ``Enhanced finite-wave-vector conductivity at multiple even-denominator
  filling factors in two-dimensional electron systems,''
  \href{http://dx.doi.org/10.1103/PhysRevB.47.7344}{{\em Phys. Rev. B}
  {\bfseries 47} (Mar, 1993) 7344--7347}.

\bibitem{halperin1993}
B.~I. Halperin, P.~A. Lee, and N.~Read, ``Theory of the half-filled landau
  level,'' \href{http://dx.doi.org/10.1103/PhysRevB.47.7312}{{\em Phys. Rev. B}
  {\bfseries 47} (Mar, 1993) 7312--7343}.
  \url{http://link.aps.org/doi/10.1103/PhysRevB.47.7312}.

\bibitem{kalmeyer1992}
V.~Kalmeyer and S.-C. Zhang, ``Metallic phase of the quantum hall system at
  even-denominator filling fractions,''
  \href{http://dx.doi.org/10.1103/PhysRevB.46.9889}{{\em Phys. Rev. B}
  {\bfseries 46} (Oct, 1992) 9889--9892}.
  \url{http://link.aps.org/doi/10.1103/PhysRevB.46.9889}.

\bibitem{willett1987}
R.~Willett, J.~P. Eisenstein, H.~L. St\"ormer, D.~C. Tsui, A.~C. Gossard, and
  J.~H. English, ``Observation of an even-denominator quantum number in the
  fractional quantum hall effect,''
  \href{http://dx.doi.org/10.1103/PhysRevLett.59.1776}{{\em Phys. Rev. Lett.}
  {\bfseries 59} (Oct, 1987) 1776--1779}.

\bibitem{Shahar1995}
D.~Shahar, D.~C. Tsui, M.~Shayegan, R.~N. Bhatt, and J.~E. Cunningham,
  ``Universal conductivity at the quantum hall liquid to insulator
  transition,'' \href{http://dx.doi.org/10.1103/PhysRevLett.74.4511}{{\em Phys.
  Rev. Lett.} {\bfseries 74} (May, 1995) 4511--4514}.
  \url{http://link.aps.org/doi/10.1103/PhysRevLett.74.4511}.

\bibitem{Shahar1996}
D.~Shahar, D.~C. Tsui, M.~Shayegan, E.~Shimshoni, and S.~L. Sondhi, ``Evidence
  for charge-flux duality near the quantum hall liquid-to-insulator
  transition,'' \href{http://dx.doi.org/10.1126/science.274.5287.589}{{\em
  Science} {\bfseries 274} no.~5287, (1996) 589--592}.

\bibitem{Wong1996}
L.~W. Wong, H.~W. Jiang, and W.~J. Schaff, ``Universality and phase diagram
  around half-filled landau levels,''
  \href{http://dx.doi.org/10.1103/PhysRevB.54.R17323}{{\em Phys. Rev. B}
  {\bfseries 54} (Dec, 1996) R17323--R17326}.
  \url{http://link.aps.org/doi/10.1103/PhysRevB.54.R17323}.

\bibitem{Note1}
Our statement regarding the observation of particle-hole symmetry at
  half-filling assumes adiabatic continuity of the measurements of strongly
  disordered samples to those with less disorder.

\bibitem{kivelson1997}
S.~A. Kivelson, D.-H. Lee, Y.~Krotov, and J.~Gan, ``{Composite-fermion Hall
  conductance at $\nu = 1/2$},''
  \href{http://dx.doi.org/10.1103/PhysRevB.55.15552}{{\em Phys. Rev. B}
  {\bfseries 55} (Jun, 1997) 15552--15561}.
  \url{http://link.aps.org/doi/10.1103/PhysRevB.55.15552}.

\bibitem{Note2}
An alternative possibility requires the composite fermion resistivity $\rho
  _{xx}^{(cf)}$ to be vanishingly small, however, in experimental systems, the
  measured $\rho _{xx} = \rho _{xx}^{(cf)} \sim (1/10 - 1) {h \over e^2}$,
  thereby contradicting the approximate PH symmetry.

\bibitem{Kamburov2014}
D.~Kamburov, Y.~Liu, M.~A. Mueed, M.~Shayegan, L.~N. Pfeiffer, K.~W. West, and
  K.~W. Baldwin, ``What determines the fermi wave vector of composite
  fermions?,'' \href{http://dx.doi.org/10.1103/PhysRevLett.113.196801}{{\em
  Phys. Rev. Lett.} {\bfseries 113} (Nov, 2014) 196801}.
  \url{http://link.aps.org/doi/10.1103/PhysRevLett.113.196801}.

\bibitem{KamburovMueed2014}
D.~Kamburov, M.~A. Mueed, I.~Jo, Y.~Liu, M.~Shayegan, L.~N. Pfeiffer, K.~W.
  West, K.~W. Baldwin, J.~J.~D. Lee, and R.~Winkler, ``Determination of fermi
  contour and spin polarization of $\ensuremath{\nu}=\frac{3}{2}$ composite
  fermions via ballistic commensurability measurements,''
  \href{http://dx.doi.org/10.1103/PhysRevB.90.235108}{{\em Phys. Rev. B}
  {\bfseries 90} (Dec, 2014) 235108}.
  \url{http://link.aps.org/doi/10.1103/PhysRevB.90.235108}.

\bibitem{LiuDengWaltz}
Y.~Liu, H.~Deng, M.~Shayegan, L.~N. Pfeiffer, K.~W. West, and K.~W. Baldwin,
  ``Composite fermions waltz to the tune of a wigner crystal.''
  arXiv:1410.3435.

\bibitem{BMF2015}
M.~Barkeshli, M.~Mulligan, and M.~P.~A. Fisher, ``Particle-hole symmetry and
  the composite fermi liquid,''
  \href{http://dx.doi.org/10.1103/PhysRevB.92.165125}{{\em Phys. Rev. B}
  {\bfseries 92} (Oct, 2015) 165125}.
  \url{http://link.aps.org/doi/10.1103/PhysRevB.92.165125}.

\bibitem{Note3}
See \cite {BalramRifmmodeCsabaJain2015} for an alternative interpretation of
  the experiments in \cite {Kamburov2014, KamburovMueed2014,LiuDengWaltz}.

\bibitem{MrossEssinAlicea2015}
D.~F. Mross, A.~Essin, and J.~Alicea, ``Composite dirac liquids: Parent states
  for symmetric surface topological order,''
  \href{http://dx.doi.org/10.1103/PhysRevX.5.011011}{{\em Phys. Rev. X}
  {\bfseries 5} (Feb, 2015) 011011}.
  \url{http://link.aps.org/doi/10.1103/PhysRevX.5.011011}.

\bibitem{Son2015}
D.~T. Son, ``Is the composite fermion a dirac particle?,''
  \href{http://dx.doi.org/10.1103/PhysRevX.5.031027}{{\em Phys. Rev. X}
  {\bfseries 5} (Sep, 2015) 031027}.
  \url{http://link.aps.org/doi/10.1103/PhysRevX.5.031027}.

\bibitem{WangSenthilfirst2015}
C.~Wang and T.~Senthil, ``Dual dirac liquid on the surface of the electron
  topological insulator,''
  \href{http://dx.doi.org/10.1103/PhysRevX.5.041031}{{\em Phys. Rev. X}
  {\bfseries 5} (Nov, 2015) 041031}.
  \url{http://link.aps.org/doi/10.1103/PhysRevX.5.041031}.

\bibitem{maxashvin2015}
M.~A. Metlitski and A.~Vishwanath, ``Particle-vortex duality of 2d dirac
  fermion from electric-magnetic duality of 3d topological insulators.'' arXiv:
  1505.05142.

\bibitem{KMTW2015}
S.~Kachru, M.~Mulligan, G.~Torroba, and H.~Wang, ``Mirror symmetry and the
  half-filled landau level,''
  \href{http://dx.doi.org/10.1103/PhysRevB.92.235105}{{\em Phys. Rev. B}
  {\bfseries 92} (Dec, 2015) 235105}.
  \url{http://link.aps.org/doi/10.1103/PhysRevB.92.235105}.

\bibitem{WangSenthilsecond2016}
C.~Wang and T.~Senthil, ``Half-filled landau level, topological insulator
  surfaces, and three-dimensional quantum spin liquids,''
  \href{http://dx.doi.org/10.1103/PhysRevB.93.085110}{{\em Phys. Rev. B}
  {\bfseries 93} (Feb, 2016) 085110}.
  \url{http://link.aps.org/doi/10.1103/PhysRevB.93.085110}.

\bibitem{Geraedtsetal2015}
S.~D. Geraedts, M.~P. Zaletel, R.~S.~K. Mong, M.~A. Metlitski, A.~Vishwanath,
  and O.~I. Motrunich, ``The half-filled landau level: the case for dirac
  composite fermions.'' arXiv: 1508.04140.

\bibitem{MurthyShankar2016halfull}
G.~Murthy and R.~Shankar, ``$\ensuremath{\nu}=\frac{1}{2}$ landau level:
  Half-empty versus half-full,''
  \href{http://dx.doi.org/10.1103/PhysRevB.93.085405}{{\em Phys. Rev. B}
  {\bfseries 93} (Feb, 2016) 085405}.
  \url{http://link.aps.org/doi/10.1103/PhysRevB.93.085405}.

\bibitem{mrossaliceamotrunich2015}
D.~F. Mross, J.~Alicea, and O.~I. Motrunich, ``Explicit derivation of duality
  between a free dirac cone and quantum electrodynamics in (2+1) dimensions.''
  arXiv:1510.08455.

\bibitem{simon1998}
S.~H. Simon, ``The chern-simons fermi liquid description of fractional quantum
  hall states,'' in {\em Composite Fermions}, O.~Heinonen, ed.
\newblock World Scientific, Singapore, 1998.
\newblock \href{http://arxiv.org/abs/arXiv:cond-mat/9812186}{{\ttfamily
  arXiv:cond-mat/9812186}}.

\bibitem{Fradkinbook}
E.~Fradkin, {\em {Field Theories of Condensed Matter Physics}}.
\newblock Cambridge University Press, 2013.

\bibitem{Note4}
In terms of the charge-conjugation ${\protect \cal C}$ and time-reversal
  ${\protect \cal T}$ transformations, PH acts by ${\protect \cal T}$ on the
  emergent gauge fields and coordinates, and by ${\protect \cal CT}$ on the
  electromagnetic gauge field.

\bibitem{Note5}
This shift is reminiscent of the shift of the three spatial dimensional quantum
  electrodynamic effective lagrangian by a term proportional $\epsilon _{\mu
  \nu \rho \tau } \partial _\mu A_{\nu } F^{\rho \tau }$ due to the (anomalous)
  transformation of the fermionic path integral measure under a chiral
  rotation.

\bibitem{Elitzur89}
S.~Elitzur, G.~W. Moore, A.~Schwimmer, and N.~Seiberg, ``{Remarks on the
  Canonical Quantization of the Chern-Simons-Witten Theory},''
\href{http://dx.doi.org/10.1016/0550-3213(89)90436-7}{{\em Nucl. Phys. B}
  {\bfseries 326} (1989) 108}.

\bibitem{wengaplessboundary}
X.~G. Wen, ``Gapless boundary excitations in the quantum hall states and in the
  chiral spin states,'' {\em Phys. Rev B} {\bfseries 43} no.~13, (1991) 11025.

\bibitem{StoneIW}
M.~Stone, ``Edge waves in the quantum hall effect,''
\href{http://dx.doi.org/10.1016/0003-4916(91)90177-A}{{\em Annals Phys.}
  {\bfseries 207} (1991) 38--52}.

\bibitem{Note6}
The assumption of non-interacting fermions is inessential and is merely made to
  simplify the description. Weak interactions may be included using
  bosonization.

\bibitem{Note7}
In the linearized limit, such perturbations are identified as the unique
  single-particle interaction that scatters a left-moving fermion into a
  right-moving fermion on the wire above it while maintaining translations
  along the $x$-direction and charge conservation.

\bibitem{Note8}
It is generally an interesting, unsolved question to determine the degree to
  which disorder is important to the stabilization of the metallic behavior
  observed about half-filing in the two-dimensional electron gas. We do not
  consider the role played by weak (or strong, for that matter) disorder in
  this paper, however, such considerations are well worth further study.

\bibitem{mulliganraghu2015}
M.~Mulligan and S.~Raghu, ``Composite fermions and the field-tuned
  superconductor-insulator transition.'' arXiv:1509.07865.

\bibitem{Hebard1990}
A.~F. Hebard and M.~A. Paalanen, ``Magnetic-field-tuned
  superconductor-insulator transition in two-dimensional films,''
  \href{http://dx.doi.org/10.1103/PhysRevLett.65.927}{{\em Phys. Rev. Lett.}
  {\bfseries 65} (Aug, 1990) 927--930}.
  \url{http://link.aps.org/doi/10.1103/PhysRevLett.65.927}.

\bibitem{MasonKapitulnik1999}
N.~Mason and A.~Kapitulnik, ``Dissipation effects on the
  superconductor-insulator transition in 2d superconductors,''
  \href{http://dx.doi.org/10.1103/PhysRevLett.82.5341}{{\em Phys. Rev. Lett.}
  {\bfseries 82} (Jun, 1999) 5341--5344}.
  \url{http://link.aps.org/doi/10.1103/PhysRevLett.82.5341}.

\bibitem{Kapitulnik2001}
A.~Kapitulnik, N.~Mason, S.~A. Kivelson, and S.~Chakravarty, ``Effects of
  dissipation on quantum phase transitions,''
  \href{http://dx.doi.org/10.1103/PhysRevB.63.125322}{{\em Phys. Rev. B}
  {\bfseries 63} (Mar, 2001) 125322}.
  \url{http://link.aps.org/doi/10.1103/PhysRevB.63.125322}.

\bibitem{Yazdani1995}
A.~Yazdani and A.~Kapitulnik, ``Superconducting-insulating transition in
  two-dimensional $\mathit{a}$-moge thin films,''
  \href{http://dx.doi.org/10.1103/PhysRevLett.74.3037}{{\em Phys. Rev. Lett.}
  {\bfseries 74} (Apr, 1995) 3037--3040}.
  \url{http://link.aps.org/doi/10.1103/PhysRevLett.74.3037}.

\bibitem{Breznay2015}
N.~P. {Breznay}, M.~A. {Steiner}, S.~A. {Kivelson}, and A.~{Kapitulnik},
  ``{Self Duality and a possible ''Hall Insulator'' phase near the
  Superconductor to Insulator Transition in two-dimensional indium-oxide
  films},'' {\em ArXiv e-prints} (Apr., 2015) ,
  \href{http://arxiv.org/abs/1504.08115}{{\ttfamily arXiv:1504.08115
  [cond-mat.supr-con]}}.

\bibitem{Moore1991}
G.~Moore and N.~Read, ``Nonabelions in the fractional quantum hall effect,''
  \href{http://dx.doi.org/http://dx.doi.org/10.1016/0550-3213(91)90407-O}{{\em
  Nuclear Physics B} {\bfseries 360} no.~2--3, (1991) 362 -- 396}.
  \url{http://www.sciencedirect.com/science/article/pii/055032139190407O}.

\bibitem{lee2007antipf}
S.-S. Lee, S.~Ryu, C.~Nayak, and M.~P.~A. Fisher, ``{Particle-Hole Symmetry and
  the $\nu={5 \over 2}$ Quantum Hall State},''
  \href{http://dx.doi.org/10.1103/PhysRevLett.99.236807}{{\em Phys. Rev. Lett.}
  {\bfseries 99} (Dec, 2007) 236807}.
  \url{http://link.aps.org/doi/10.1103/PhysRevLett.99.236807}.

\bibitem{levin2007}
M.~Levin, B.~I. Halperin, and B.~Rosenow, ``Particle-hole symmetry and the
  pfaffian state,'' \href{http://dx.doi.org/10.1103/PhysRevLett.99.236806}{{\em
  Phys. Rev. Lett.} {\bfseries 99} (Dec, 2007) 236806}.
  \url{http://link.aps.org/doi/10.1103/PhysRevLett.99.236806}.

\bibitem{CanoHughesMulligan2015}
J.~Cano, T.~L. Hughes, and M.~Mulligan, ``Interactions along an entanglement
  cut in $2+1\mathrm{D}$ abelian topological phases,''
  \href{http://dx.doi.org/10.1103/PhysRevB.92.075104}{{\em Phys. Rev. B}
  {\bfseries 92} (Aug, 2015) 075104}.
  \url{http://link.aps.org/doi/10.1103/PhysRevB.92.075104}.

\bibitem{BisharaNayak}
W.~Bishara and C.~Nayak, ``Effect of landau level mixing on the effective
  interaction between electrons in the fractional quantum hall regime,''
  \href{http://dx.doi.org/10.1103/PhysRevB.80.121302}{{\em Phys. Rev. B}
  {\bfseries 80} (Sep, 2009) 121302}.
  \url{http://link.aps.org/doi/10.1103/PhysRevB.80.121302}.

\bibitem{SpivakKivelsonemulsions}
B.~Spivak and S.~Kivelson, ``Transport in two dimensional electronic
  mirco-emulstions,'' {\em Annals of Physics} {\bfseries 321} (2006) 2071.

\bibitem{JameiKivelsonSpivak2005}
R.~Jamei, S.~Kivelson, and B.~Spivak, ``Universal aspects of coulomb-frustrated
  phase separation,''
  \href{http://dx.doi.org/10.1103/PhysRevLett.94.056805}{{\em Phys. Rev. Lett.}
  {\bfseries 94} (Feb, 2005) 056805}.
  \url{http://link.aps.org/doi/10.1103/PhysRevLett.94.056805}.

\bibitem{BalramRifmmodeCsabaJain2015}
A.~C. Balram, C.~T\ifmmode~\mbox{\H{o}}\else \H{o}\fi{}ke, and J.~K. Jain,
  ``Luttinger theorem for the strongly correlated fermi liquid of composite
  fermions,'' \href{http://dx.doi.org/10.1103/PhysRevLett.115.186805}{{\em
  Phys. Rev. Lett.} {\bfseries 115} (Oct, 2015) 186805}.
  \url{http://link.aps.org/doi/10.1103/PhysRevLett.115.186805}.

\end{thebibliography}\endgroup
\bibliographystyle{utphys}

\end{document}